\documentclass[11pt]{book}
\usepackage{Wiley-AuthoringTemplate}
\usepackage{chapterbib}
\usepackage[sectionbib,numbers]{natbib}

\usepackage{amsmath,amssymb,amsthm}
\usepackage{amssymb,amsmath,amsfonts,eurosym,geometry,ulem,graphicx,color,setspace,sectsty,comment,footmisc,pdflscape,subfigure,array}
\usepackage{multirow}
\usepackage{enumitem}
\usepackage{tikz,bm}
\usepackage{mathtools}
\usepackage{diagbox}
\usepackage{algpseudocode}
\usepackage{color, colortbl}
\usepackage{tabularx}
\usepackage{dsfont}
\definecolor{Gray}{gray}{0.9}
\usepackage{tcolorbox}
\usepackage{centernot}
\usepackage{float,bbm}

\usepackage[ruled,vlined]{algorithm2e}
\usepackage{algpseudocode}

\makeindex

\begin{document}


\mainmatter
\include{p01}

\chapter[Zero-Trust Cyber Resilience]{Zero-Trust Cyber Resilience\protect}

\author*[1]{Yunfei Ge}
\author[1]{Quanyan Zhu}

\address[1]{\orgdiv{Department of Electrical and Computer Engineering}, 
\orgname{New York University}, 
\postcode{11201}, \countrypart{New York}, 
     \city{Brooklyn}, \street{370 Jay Street}, \country{USA}}%

\address*{Corresponding Author: \email{yg2047@nyu.edu}}

\maketitle

\begin{abstract}{Abstract}

The increased connectivity and potential insider threats make traditional network defense vulnerable. Instead of assuming that everything behind the security perimeter is safe, the zero-trust security model verifies every incoming request before granting access. This chapter draws attention to the cyber resilience within the zero-trust model. We introduce the evolution from traditional perimeter-based security to zero trust and discuss their difference. Two key elements of the zero-trust engine are trust evaluation (TE) and policy engine (PE). We introduce the design of the two components and discuss how their interplay would contribute to cyber resilience.
Dynamic game theory and learning are applied as quantitative approaches to achieve automated zero-trust cyber resilience.  Several case studies and implementations are introduced to illustrate the benefits of such a security model.

\end{abstract}


\section{Introduction}\label{sec:introduction}

The world is increasingly connected with the recent advances in cloud services, data communications, and automation technologies. While such technologies increase flexibility and efficiency, the adoption of smart devices and the Internet of Things (IoT) has brought with it new and expanding cyber risks that have the potential to impact not just a particular entity or industry but are a serious concern for all private and public industries alike \cite{chen2019optimal}. 
First, the increased connectivity inevitably enlarges the attack surface and enables the attacker to access the system from multiple entry points. 
Second, modern networks, especially massive IoT networks, consist of heterogeneous devices, diverse applications, and third-party services, and not all of them are accompanied by regular security updates \citep{zhu2021cybersecurity}. 
Moreover, recent trends such as remote work and bring your own device (BYOD) exacerbate the need for a trustworthy environment to provide proper access and authentication mechanisms for external connections from unknown or uncertain environments \citep{weil2020risk}. 
It becomes a challenging task to defend against all vulnerabilities and manage the security of IoT networks as their size and coverage grow.

\begin{figure}[!t]{
	\begin{minipage}[c][1\width]{
	   0.45\textwidth}
	   \centering
	   \includegraphics[width=1\textwidth]{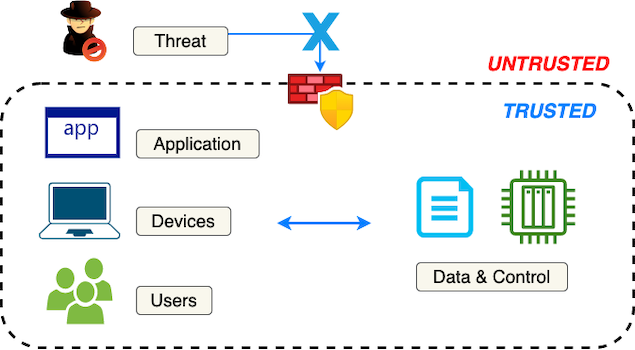}
	    \caption{Traditional security perimeter architecture assumes everything inside the perimeter is trusted.}
	    \label{fig:trad}
	\end{minipage}}
 \hfill 	
{
	\begin{minipage}[c][1\width]{
	   0.45\textwidth}
	   \centering
	   \includegraphics[width=1\textwidth]{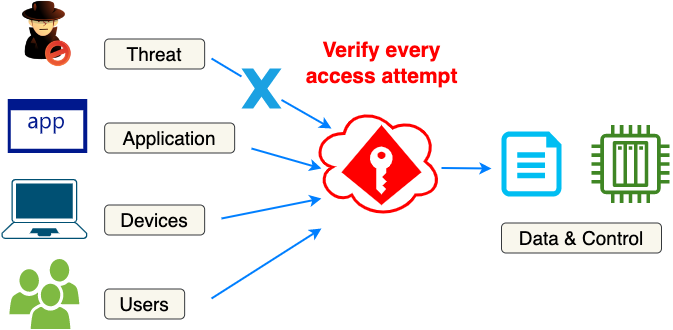}
	    \caption{Zero-trust security architecture verifies every access attempt, including users, devices, and applications.}
	     \label{fig:ztgeneral}
	\end{minipage}}
\end{figure}

Traditional perimeter-based defense, e.g., firewalls, intrusion detection, systems (IDS), and virtual private networks (VPN), aims to keep the attacker outside the security perimeter. As illustrated in Figure~\ref{fig:trad}, the traditional security architecture divides the networks into trusted and untrusted zones by establishing the defensible boundary between internal assets and the outside world. However, they become insufficient to defend against sophisticated adversaries, including insider threats \cite{huang2021duplicity} and advanced persistent threats (APTs) \cite{zhu2018multi}. In these attack scenarios, the attacker can evade traditional protections, obtain privileges as an insider with stolen credentials, and move laterally in the network toward the primary target. Once attackers breach the perimeter, they will be inherently trusted and unhindered to achieve their goals. 
Modern networks must transform from static and perimeter-based defenses to a strategic and dynamic trust-based security framework that focuses on the identity and integrity of individual components in the network at present.

The complexity in network structure and the sophistication of the attack model lead to an emerging security solution, \textbf{Zero trust} (ZT).
Zero trust is a new security concept that forfeits the assumption that everything behind the security perimeter is safe \cite{stafford2020zero}. With the principle ``never trust, always verify'', zero trust eliminates implicit trust and continuously validates the identity and integrity of each entity. Regardless of the physical location (inside or outside the network), any request for access to critical data or services would first be verified through the zero-trust engine, locally distributed or at the cloud, before granting access. The zero-trust engine would verify the user, device, and application associated with the request based on fine-grained access policies. Once the zero-trust engine decides the access attempt will be approved, it dynamically updates the network configuration to accept the request. It aims to prevent data breaches and limit internal lateral movement.

Zero trust plays a major role in building cyber resilience, as it provides additional layers of protection to withstand unpredictable threats and limit impact if an unauthorized entity does gain access. It enables fine-grained security policies based on the trustworthiness of the network component. With accurate trust evaluation, the system can quickly identify suspicious components in the system, implement proper defense against potential threats, and efficiently respond to cyber incidents after the attack. Besides, it can incentivize compliant behaviors of the network users and reduce the temptation for insiders to abuse or misuse the network \cite{colwill2009human}, as they need to be sufficiently trustworthy in order to accomplish their missions. The security posture of the system can be more efficient, visible, compliant, and cost-efficient with trust-based defense.

However, several challenges arise from the design of zero-trust security models. First, there is a need to quantitatively define and measure the trustworthiness of the agent so that metrics can be used for planning and policy design.
Second, in highly connected networks, constant monitoring and implementing defense with maximum security at all times can cause a time delay and degrade the performance of the system \cite{zhu2009dynamic}. Hence, strategic zero trust needs to be employed for the sake of balancing system performance and security. Third, the mobility and the changing topology of IoT networks create a dynamic environment. Based on the baseline defense policy, the security decisions also need to accommodate the environmental change promptly and craft the policies adaptively with online learning.

In this chapter, we introduce a quantitative design framework to provide a formal design methodology to address these challenges in zero-trust design. We propose the strategic zero-trust principle:\textbf{\textit{``Never trust, verify (defense) strategically''}}. To formalize the zero-trust design, game theory has been a natural and successful framework for modeling the interactions between the system and the potential attacker. Many problems in cyber security are fundamentally decision-making problems under complex, uncertain, multi-agent conditions \citep{kamhoua2021game}. The game theory equilibrium analysis not only provides a way to quantitatively assess the security posture of the network but also enables a formal methodology to design best-effort zero-trust policies. The optimal security policy aims to maximize the non-myopic well-being of the system by looking ahead multiple steps into the horizon. We hope that by using a game-theoretic approach, the system can achieve automated zero-trust security with better efficiency.

The chapter is organized as follows. Section~\ref{sec:trad} introduces the general framework of zero-trust defense and compares it with traditional perimeter-based defense. We elaborate on the design of two zero trust core components, trust evaluation and policy engine, in Section~\ref{sec:trust} and Section~\ref{sec:policy}. Section~\ref{sec:resi} discusses how zero trust contributes to cyber resilience. Then, we propose our strategic zero trust design in Section~\ref{sec:case} and introduce several works that apply game theory to cyber security. Section~\ref{sec:conclusion} concludes the chapter.

\section{From Traditional Security to Zero Trust}
\label{sec:trad}

Zero-trust security is a natural evolution of traditional perimeter-based security. It pushes the trust boundary down to every stage of digital interaction, offering granular access control and better security. A comprehensive zero-trust approach encompasses not only users but also devices and applications inside the network. For users, the system should apply strong authorization and authentication processes to ensure user identity. For devices (e.g., routers, controllers, end-point devices, etc.), the system should ensure the devices are securely managed. For applications, a zero-trust system needs to remove implicit trust and enforce appropriate in-app permissions. Considerable efforts have been invested in research and implementation of zero-trust security. The National Institute of Standards and Technology (NIST) report \cite{stafford2020zero} outlines the basic architecture and deployment of the zero-trust model. It has been applied to enterprise security (e.g., Google BeyondCrop \cite{beyondcorp}), cloud computing security \cite{li2012cyberguarder}, big data \cite{tao2018fine}, etc.

Zero Trust takes a holistic view of network security and establishes access policies based on the trustworthiness of each network component. The goal of zero trust is to ensure every stage of digital data flow is trusted. The key question is to answer who is asking for what under which condition. Only when all the entities involved are trusted,  the zero trust engine can allow permission for the request. To achieve this, continuous dynamic access control is needed, which should not only depend on the IP address of the request but also on the historical behavior and real-time analysis of the components.

\begin{figure}[ht!]
 \centering
 \includegraphics[width=0.9\textwidth]{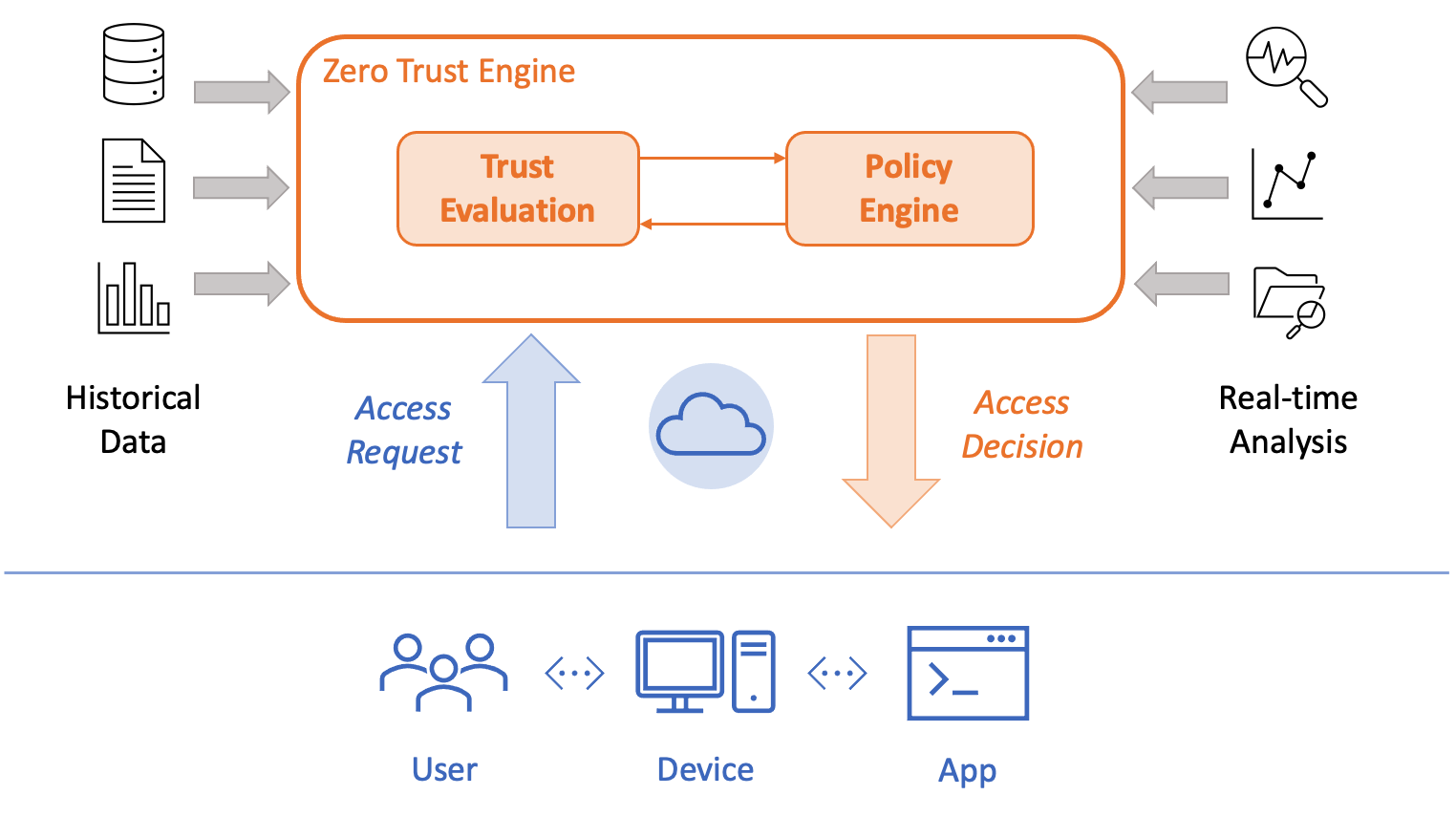}
 \caption{General zero trust workflow. Every access request is sent to the zero-trust engine for approval. The engine will evaluate the trustworthiness of the involved entities based on historical data and real-time analysis. The policy engine makes access decisions based on the trust evaluation. }
 \label{fig:ztwork}
\end{figure}

The general zero-trust workflow is illustrated in Figure~\ref{fig:ztwork}. The zero-trust engine is mainly composed of two parts: trust evaluation (TE) and policy engine (PE). When receiving an access request, the zero-trust engine first evaluates the trustworthiness of the involved entities at TE based on historical data and observed footprints through continuous monitoring information. Given the evaluation result, PE makes access decisions for the request and enforces different policies to network resources. 

Zero trust is never a single technology or application, but a framework with a set of principles that guides the design of the security architecture in the network \cite{ge2022mufaza}.  Zero trust security is built upon existing network management tools and seeks to address modern cyber threats in novel ways. It takes into account the trustworthiness of network components in decision-making and provides a fine-grained access control that can be adapted to varying conditions. 

Table~\ref{tab:tradvszt} illustrate the difference between traditional perimeter-based security and zero-trust security by comparing some of the key features. It should be noted that comprehensive zero-trust protection brings security, but is also accompanied by problems of higher implementation cost and network complexity. Achieving a balance between security and cost is an important consideration for any zero-trust system.

\begin{table}[ht!]
\caption{Comparison between traditional and zero-trust security.\label{tab:tradvszt}}{%
\begin{tabular}{@{}|c|c|c|@{}}
\toprule
  \rowcolor{Gray}
\textbf{Features}     & \textbf{Traditional}   &\textbf{ Zero Trust }\\
\midrule
Principle & Trust but verify & Never trust, always verify\\\hline
Trust Boundary & Entire infrastructure & Individual component\\\hline
Security Consideration & Device and location  & User, device, and application\\\hline
Access Control & IP-based access control & Trust-based access control\\\hline
Authentication & One-time at initial access & Continuous authentication\\\hline
Implementation Cost & Low & High\\
\botrule
\end{tabular}}{}
\end{table}

\section{Trust Evaluation (TE) Design}
\label{sec:trust}

Trust is the key component affecting zero-trust security policies. A wide variety of literature has studied trust in different disciplines, including psychology, economics, political science, sociology, and recently computer science. 
In general, trust is a measure of confidence that an entity will behave expectedly. 
The trust-based decision will only be considered when there is a chance of deception by the opponent. In cyber security, this refers to cases when the attacker deliberately misleads or conceals information in order to achieve his strategic advantage. It is important to understand the definition, metrics, and evaluation methods of trust to design an efficient TE in a zero-trust framework. We summarize the necessary attributes to design TE in computer networks in Table~\ref{tab:trust}.

\begin{table}[ht!]
\caption{Trust definition attributes in computer networks.\label{tab:trust}}{%
\begin{tabular}{@{}|c|c|@{}}
\toprule
  \rowcolor{Gray}
\textbf{Attributes}     & \textbf{Explanation}   \\
\midrule
Target  & Who is the entity that will be evaluated\\\hline
Metric & What is the metric that is used to measure trust\\\hline
Collection & What information is collected to calculate trust\\\hline
Evaluation & How to evaluate trust\\\hline
Purpose  & How trust will be used in decision-making\\\hline
Management & How to manage the trust information in the system\\\hline
\botrule
\end{tabular}}{}
\end{table}

\subsection{Target} 
Different from transitional perimeter-based security, zero trust expands the target of trust to every component in the network. Trust decisions will be based on not only the trustworthiness of the requiter but also on the device and environment where the data flow takes place. The granularity of the target in TE defense depends on the computational capability and the need of the system. 

\subsection{Metric} 
TE adopts a metric to measure the trustworthiness of the entity and provide risk analysis for policy decisions. In this chapter, we refer to this metric as the trust score (TS). To be specific, we formalize the trust score of an entity at the current time as the probability that the entity is non-adversarial to the system. Let $\theta\in\Theta$ be the attributes of the entity $i$, and denote the non-adversarial attributes set as $\Theta_T$. Formally,
\begin{definition}[\textbf{Trust Score}]
The Trust Score (TS) of the entity $i$ at time $t$ is defined as the probability that the entity is non-adversarial to the system:
\begin{align}
    TS^t(i):= \Pr(\theta^t_i\in\Theta_T) \in [0,1],
    \label{eq:ts}
\end{align}
where $\theta_i^t$ is the attributes of entity $i$ at time $t$. 
\label{def:TS}
\end{definition}
It should be noted that in practice, trust is multi-faceted and the attribute $\theta$ can be a multi-dimensional vector where each entry represents different trust attributes.

\subsection{Collection and Evaluation} 

We adopt the categories from Bonatti et al.\citep{bonatti2007integration} and discuss two common approaches to trust collection and evaluation: policy-based and reputation-based trust management. Then, we propose our approach of Bayesian trust evaluation, which is a combination of policy-based and reputation-based methods. 

\subsubsection*{Policy-based Method:}
Policy-based methods enable the system to manage trust based on a set of predefined policies. These policies may include rules that specify the types of users or devices that are allowed to access certain resources, the level of access that is granted, and the conditions under which access is granted or denied. The trust is established based on collected hard-evidences, e.g., credentials, access tokens, and certificates. The evaluation process is relatively simple as the trust engine only needs to determine whether the collected data satisfy the predefined conditions. However, it is important to tailor the security policy to the needs of the system. We provide some examples under this category.

\begin{itemize}
    \item \textbf{Network credential.}  The access request can be granted based on the given credentials of the entity. The trust information of the entity is encrypted in the credential as we assume only the trusted entity will process the credential. Kerberos \citep{neuman1994kerberos} is one example of authenticating service requests between trusted hosts across an untrusted network, such as the internet. The underlying requirement for this method is that the system needs to ensure that the credential is private and not revealed to the attacker.
    
    \item \textbf{Ad-hoc attributes check.} The system can configure a set of qualified attributes that must be met before access is allowed. These attributes may include the device configuration, network environment security, application permission, etc. Identifying the necessary security attributes requires extensive knowledge of the system vulnerabilities. Poor security checks could result in inaccurate estimation of TS along with unresolved security vulnerabilities.

    \item \textbf{Promise and incentive compliance.} Trust can also be influenced by promises and penalties. The system can develop a set of rules or contracts that encourages incentive-compatible behaviors of the agent. The system can strategically design a reward and penalty mechanism to elicit desirable safe behaviors of the entity. In particular, the integration of cyber deception (e.g., honeypots, obfuscation \citep{pawlick2021game}, and incentive modulation \citep{huang2021duplicity}) into IoT systems provide a proactive way to detect and respond to APT attackers.
\end{itemize}

\subsubsection*{Reputation-based Method:}

Reputation-based methods estimate the trustworthiness of an entity and adjust access permissions based on interactions or observations from past experiences, either directly (historical behaviors) or indirectly (third-party recommendation). What information we need to collect depends on the availability of corresponding data and the needs of the system. This method can integrate more information but is also more vulnerable to false positives or false negatives. The system needs to consider the reliability of the source of reputation during trust evaluation. We provide some examples under this category.

\begin{itemize}
    \item \textbf{Historical behaviors.} If the system has had direct interactions with the entity, the TS of the entity at this request can be developed based on the history of their encounters. The behavior characteristics of the entity can be multi-dimensional that involve login information, operational habits, abnormal behavior record, etc. The system needs to find proper risk measures that achieve the security goals. In addition, expired experiences should be excluded from the trust evaluation due to the dynamic features of modern networks. The out-of-date interaction record contributes little to the current trustworthiness of the entity and it is important to consider the attenuation in the data history.

    \item \textbf{Social Reputation.} Reputation from third parties or society can also serve as a source for trust evaluation.  Reputation may be defined as the global perception of the entity as being trustworthy. In other words, it is a collective trust opinion of other systems about the behavior of a subject node. The system would prefer to grant access to a well-reputed entity. This information would be helpful when the entity is trying to enter the system for the first time.

    \item \textbf{Recommendation.} Recommendation is the simplest case of trust propagation. For instance, a recommendation from a trusted neighbor would increase the TS of the new entity. Reliable recommendation reduces information overload, uncertainties, and risk of the access attempt. It is important to provide a trust inference model to find a reliable recommendation that could improve trust evaluation accuracy.
    
    \item \textbf{Supply Chain.} Supply chain contains inter-organizational relationships among interdependent companies contributing to the final components in the target system. The trust in the suppliers would also influence the trust evaluation of the device they provided. This type of trust propagation is multi-hop due to the multi-tier structure in the supply chain. Accountability investigation and cyber insurance in the supply chain \citep{ge2022accountability} could encourage truth-telling and information transparency to support trust evaluation.

    \item \textbf{Third-party Evaluation.} Besides the trust information propagated from others, the security system can also leverage third-party evaluation results (e.g., Intrusion Detection System (IDS) \citep{zhu2012guidex}, Security Information and Event Management (SIEM) \citep{miller2011security}, etc.) to derive a more reliable measure of trust score of the entity. The trace of the user provides a sequence of events that can be used for security analysis. It should be noted that the reliability of the side evidence will largely impact trust propagation. The system needs to incorporate reliable side evidence for an accurate trust measure.
\end{itemize}

\subsubsection*{Our Approach - Bayesian Trust Evaluation}

Under the dynamic network environment, it is important for zero-trust security to continuously adjust the TS after the initial trust evaluation. The system needs to respond to changes in trust by investigating and orchestrating responses to potential incidents. The dynamic update should take account of previous knowledge about the entity as well as currently observed behaviors. In this work, we propose a Bayesian trust model to update the trust score. This model offers a quantitative way to combine policy-based trust with reputation-based evidence and update the TS subject to the perceived strategies of the entity.

\begin{definition}[\textbf{Bayesian Trust Update}]
The Trust Score (TS) of the entity $i$ at time $t+1$ is the probability that the entity is non-adversarial ($\theta^{t+1}_i\in\Theta_T$) based on the prior knowledge, side evidence, and observed strategies of the entity:
\begin{align}
    TS^{t+1}(i)=& \Pr(\theta_i^{t+1}\in\Theta_T|a^t,e^t,\pi^t) \notag \\
    =& \frac{h(e^t|a^t,\theta^t_i\in\Theta_T)\sigma(a^t|\theta^t_i\in\Theta_T)\pi^t(\theta^t_i\in\Theta_T)}{\sum_{\hat{\theta}_i\in\Theta} h(e^t|a^t,\hat{\theta}_i)\sigma(a^t|\hat{\theta}_i)\pi^t(\hat{\theta}_i)}
    \label{eq:bts}
\end{align}
where $a^t\in\mathcal{A}$ is the observed action of the entity, $e^t\in\mathcal{E}$ is the received side evidence, and $\pi^t$ is the system's prior knowledge about the entity up to time $t$. $\sigma$ is the observed strategy of the opponent and $h$ is the evidence-generating function given by the third party. Note that the relationship between $TS$ and $\pi$ is: $TS^t(i) = \pi^t(\theta^t_i\in\Theta_T)$.
\label{def:BTS}
\end{definition}

\begin{itemize}
    \item \textbf{Prior Knowledge $\pi^t$:} 
    Prior knowledge is the ex-ante likelihood of the entity being non-adversarial before taking into consideration any new (posterior) information. This information can be collected through various sources through policy-based methods or reputation-based methods. The initial trust score $TS^0(i)=\pi^0$ is usually constructed based on some kind of experience with, or firsthand knowledge of, the other party. For instance, attributes check, historical behaviors, reputation, etc. can all contribute to the prior computation. The system could establish an initial trust estimation of the entity at $t=0$ and calculate the probability of the agent being trusted, i.e., $TS^0(i)\in [0,1]$. 
    
    \item \textbf{Side Evidence $e^t\in\mathcal{E}$:} The system could also incorporate side security evidence during trust updates. The side evidence may involve real-time network detection, system monitoring information, intelligent risk analysis, security alerts, etc. Reliable evidence helps establish a fast and accurate trust evaluation \citep{ge2022mufaza}.

    For instance, the external evidence is additional information taking binary value $e^t\in \mathcal{E}= \{0,1\}$, where $e^t=1$ indicates a security alarm, and $e^t=0$ means no alarm.  The security alarm warns the defender when the agent is more likely to be malicious. In general, the evidence is generated based on the probability that the agent with type $\theta^t$ takes an action $a^t$ at the current time $t$. Observing the evidence $e^t$, the defender can further update the trust of the agent via Bayes' rule.

    \item \textbf{Observed Strategies $\sigma(a^t|\theta^t)$: } Last but not least, the observed strategies of the agent contribute a major part to the trust updates. The strategy refers to a complete contingent plan of actions that encompasses the agent's objectives. It captures the intention of the agent and infers what the agent with a different type would like to do in the current security state. One example would be observed abnormal behaviors of the agent. If the agent attempts to access sensitive or restricted information in the system, there is a great chance that the agent has been compromised by the attacker, thus the $TS$ of the agent should be decreased. 
    The system should constantly monitor the behaviors of the agent and update or re-evaluate its trustworthiness periodically. 
\end{itemize}

\subsection{Purpose}
In the zero-trust model, the TS is used to obtain trust-based access policies. It plays a key role in establishing secure communication between different systems, networks, and individuals. Depending on the needs of the system, each situation places different requirements on trust. For instance, in data communication, the trust requirements would focus on the security level of the transmission environment. On the other hand, in supply chain security, the trust of the supplier would pay more attention to the reputation and compliant behaviors of the supplier. It is important to determine the purpose of trust evaluation and design the TE accordingly.

\subsection{Management}

Two major approaches to managing trust in cyber security are centralized and distributed trust management. Centralized trust management involves a central authority or entity that is responsible for managing and enforcing trust policies across a system or network. It is typically used in environments where there is a clear hierarchy of trust relationships. Distributed trust management, on the other hand, involves a decentralized network of entities that are responsible for managing and enforcing trust policies. In this approach, trust decisions are made based on consensus among multiple entities, rather than by a single central authority. Each entity in the network may have its own trust policies and evaluation criteria, and trust decisions are made based on the collective evaluation of these policies and criteria. 

Both centralized and distributed trust management approaches have their advantages and disadvantages. Centralized trust management can provide a clear hierarchy of trust relationships and centralized enforcement of trust policies, but it may also be vulnerable to single points of failure and may require significant resources to maintain. Distributed trust management can be more resilient and adaptable to changing trust relationships, but it may also be more difficult to manage. The choice between centralized and distributed trust management in zero trust depends on the specific needs and requirements of the system or network.


\section{Policy Engine (PE) Design}
\label{sec:policy}

\begin{figure}[!ht]
 \centering
 \includegraphics[width=0.8\textwidth]{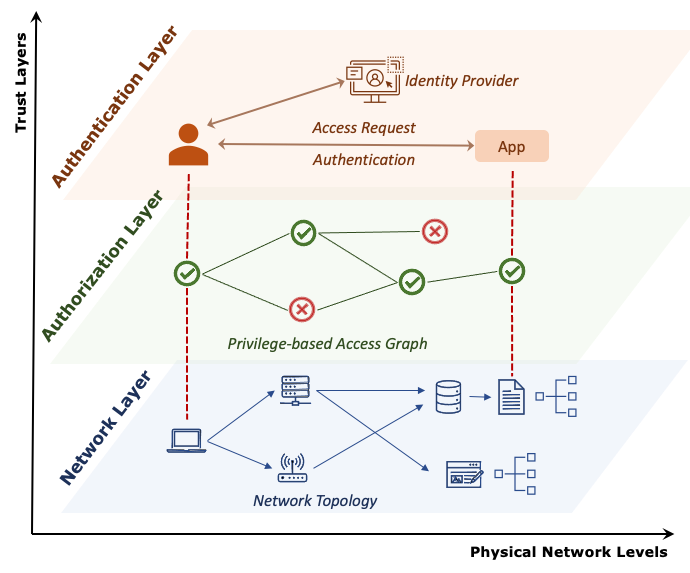}
 \caption{Multi-dimensional mitigation in zero trust security. Authentication layer ensures the user’s identity. Authorization layer determines the user’s possible access points based on his privilege. Network layer transmits the data based on the topology. The graphs on each layer are dynamic and may change over time. Single access attempt will be allowed only when it satisfies the requirements for all three layers.}
 \label{fig:trustlayer}
\end{figure}

Zero Trust is a holistic security model that provides additional layers of mitigation against sophisticated attacks. It forfeits the assumption that everything behind the security perimeter is safe, and thus requires continuous monitoring and validation of the identity and privilege associated with each network component. The key components of zero-trust security are identity verification, privilege management, and network segmentation. The fine-grained authentication and authorization process, along with network policy enforcement configuration, form comprehensive zero-trust security throughout the environment. Zero-trust architecture adds additional trust layers on top of the perimeter-based defense. It provides a multi-dimensional mitigation compared to traditional perimeter-based security. Based on this point of view, we provide our interpretation of zero trust from a trust layer perspective, as illustrated in Figure~\ref{fig:trustlayer}.

\subsection{Authentication Layer: Continuous Authentication }

The authentication layer is a critical supporting component for realizing zero trust security. It answers the question \textit{``Who you are?''} in access decision-making. Zero trust security applies identity-based instead of location-based access policies, which requires the system to conduct strict identity verification. The authentication process ensures that the attributes of the users/devices/applications coincide with the identity of who they claim to be. Under a zero-trust policy, the system continuously monitors the behaviors of the components and adjusts policies dynamically based on observations. Different authentication methods can be utilized to support this layer. The multi-factor authentication (MFA) is preferred for security as it grants access only after successfully presenting two or more pieces of evidence to the zero-trust engine. However, the system also needs to balance cyber security and system efficiency, as it would be time-consuming to conduct constant verification.

\subsection{Authorization Layer: Least Privilege Access}

This layer provides the zero-trust engine with the access graph based on the privilege level. It answers the question \textit{``Where can you go?''} before granting any access. One fundamental principle of zero-trust security is least privilege access, which grants the least amount of privilege necessary depending on who is requesting access and the context of the request. This approach mitigates the lateral movement of an attacker, who compromises an existing user account and attempts to penetrate deeper into the system. To improve policy creation and enforcement, the system needs to have a thorough knowledge of the security postures inside the network, as well as network topology and the context of the request.

\subsection{Network Layer: Micro-segmentation}

The last layer is the network layer, which configures the network topology at the software level and virtually divides the network into multiple smaller segments. It answers the question \textit{``How to get there?''} for any access request. Micro-segmentation techniques enable the system to break the network and cloud environment into secure zones at the application workload level. This approach reduces the attack surface, prevents lateral movement and data breach, and achieves regulatory compliance. It provides a method to create granular access policy and help visualize the vulnerabilities associated with applications.

In summary, zero trust enables multiple layers of defense from a holistic network system view. Traditional authentication, authorization, and network segmentation policies usually focus on IP-based local decisions and fail to consider the incentives of the access request. In zero-trust security, the policies on each trust layer of zero-trust architecture would depend on the trust evaluation and risk assessment of the request, along with the global access policy for the whole network. The access attempt will be allowed only when it satisfies the requirements for all three layers.

\section{Zero Trust for Cyber Resilience}
\label{sec:resi}

\subsection{How Zero Trust contributes to Cyber Resilience}

\begin{figure}[ht!]
 \centering
 \includegraphics[width=0.7\textwidth]{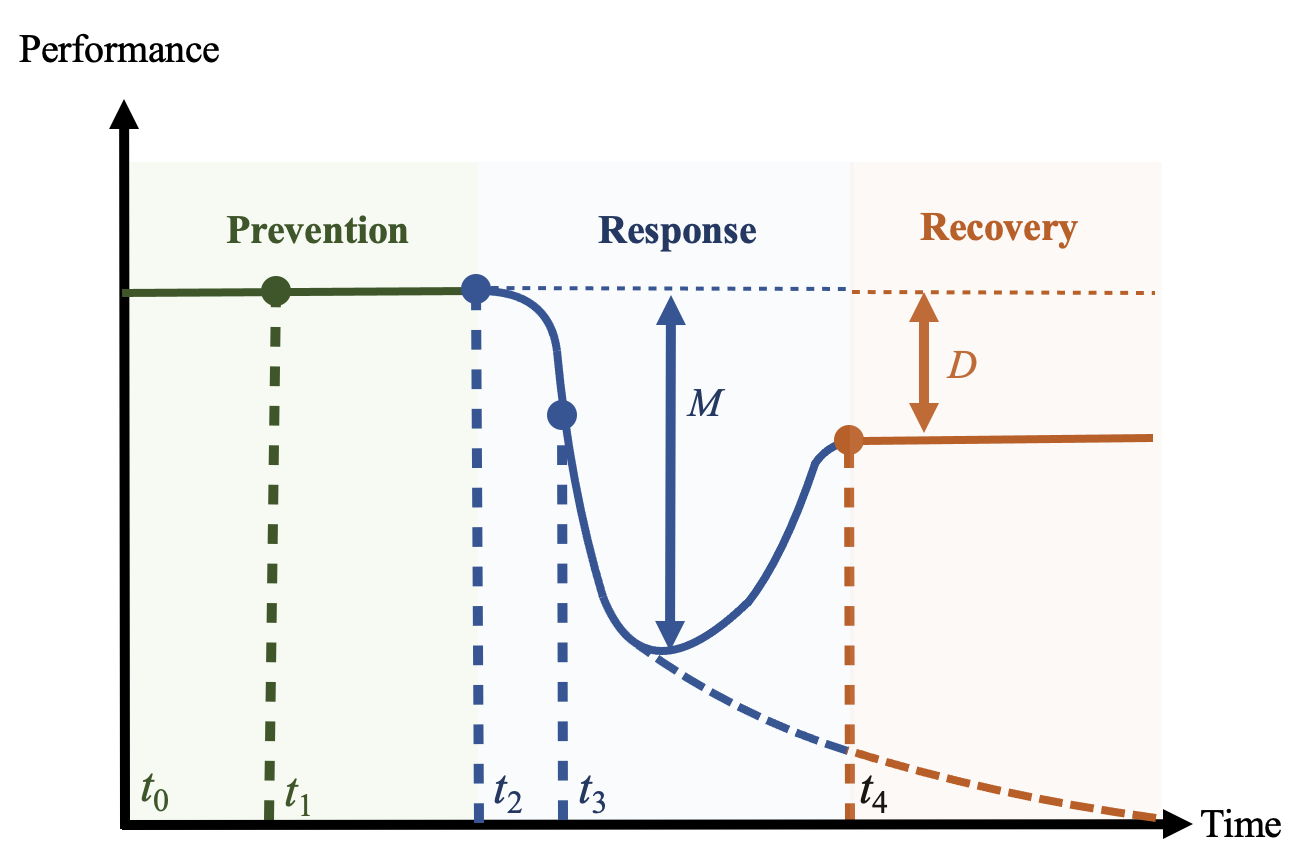}
 \caption{The resilience of the zero-trust system can be evaluated through performance evolution. Resilience includes prevention before the attack, response during the attack, and recovery after the attack. Zero trust contributes to all three stages of the resilience plot. }
 \label{fig:resilience}
\end{figure}

For a given attack, resilience is generally understood as the ability of the system to recover from an external disruptive event \cite{henry2012generic}. It involves actions before and after the incident \cite{zhu2020cross}. Figure~\ref{fig:resilience} illustrates the general performance transition when considering cyber resilience. We say the zero-trust system is $(T,D)$-resilient if the system can recover from the attack within $T=t_4-t_2$ units of time and maintain a maximum loss of performance $D$ \cite{zhao2022multi}. 

Zero trust plays a major role in building cyber resilience, as it provides additional layers of protection to prevent unpredictable threats, respond to the attack, and limit impact if an unauthorized entity does gain access. Through the section, we will explain how zero trust helps to build a resilient security ecosystem. 

\subsubsection*{Prevention}
Prevention in cyber resilience includes the ability to identify the risk and protect the system from potential threats. The first core function of cyber resilience is the ability to \textbf{identify} risks. Zero trust security forfeits the assumption that everything behind the security perimeter is safe. The continuous trust evaluation process provides visibility to the system so that organizations can identify vulnerabilities of potential threats related to users’ trustworthiness.

Where identity focuses primarily on baselining and monitoring, \textbf{protect} is when the framework becomes more proactive. The zero trust model encourages the system to proactively establish strict security controls and comprehensive incident response plans. Based on the zero-trust assumption, the systems will always be prepared for any system compromises so that they can respond to the threats effectively.

\subsubsection*{Response}
Response in cyber resilience includes the ability to detect cyber events and respond accordingly. 
The \textbf{detect} function is a critical step to a robust cyber program - the faster a cyber event is detected, the faster the repercussions can be mitigated. By continuously verifying the user’s activity, zero trust security provides organizations with greater visibility of network security \cite{de2017building}. Any changes in the user’s behavior (e.g., unusual login attempts and unauthorized access requests) would lead to a trust score update, informing the organization of any suspicious activity in the system.

Moreover, it is important to establish a \textbf{respond} function that supports the ability to contain the impact of a potential cybersecurity incident. Zero trust helps to improve cyber resilience by providing a multi-dimensional incident response to deter an attacker from a holistic system view \cite{buck2021never}. 

\subsubsection*{Recovery}
Finally, the \textbf{recover} function helps to reconfigure defense plans and restore any impaired capabilities or services after attacks. Zero trust security solutions often include automated response capabilities, which can help to quickly contain security incidents and adapt to sudden changes in the network \cite{huang2019adaptive}. We will show in a later section that game theory provides the analytical basis for establishing effective automated zero-trust security responses.

\subsection{A Running Example}
 With the aforementioned three trust layers, zero trust provides multi-dimensional mitigation against comprehensive cyber attacks and plays a major role in building cyber resilience. We provide an example to illustrate how zero-trust security helps the system build a resilient security model against insider threats and internal lateral movement at each layer.

\subsubsection*{Traditional Architecture:}

\begin{figure}[ht!]
 \centering
 \includegraphics[width=0.7\textwidth]{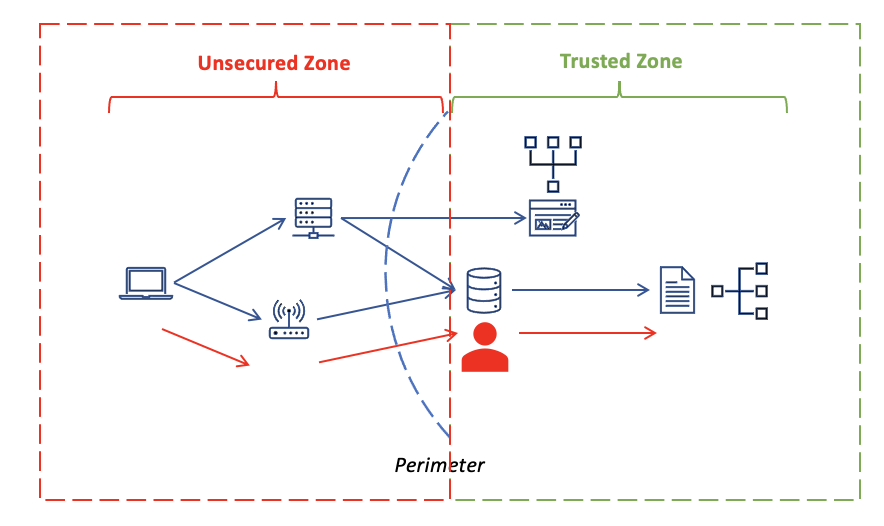}
 \caption{Attack with traditional perimeter-based defense. If the attack successfully enters the trusted zone, the system cannot stop the attacker from reaching the valuable asset using traditional perimeter-based defense.}
 \label{fig:egtrad}
\end{figure}

Consider an attacker stealthily gaining access to the victim system through phishing attacks or account compromise, as shown in Figure~\ref{fig:egtrad}.  Traditional defense tends to focus on the boundaries of the network. However, traffic within the trusted zone is usually only lightly
monitored because users inside the firewall are assumed to be trusted. With traditional perimeter-based defense, once inside the trusted zone, the attacker can disguise as an insider with stolen credentials, and move laterally in the network toward the primary target. This one-dimensional protection is inadequate when dealing with sophisticated attacks such as APTs.

\begin{figure}[ht!]
 \centering
 \includegraphics[width=0.7\textwidth]{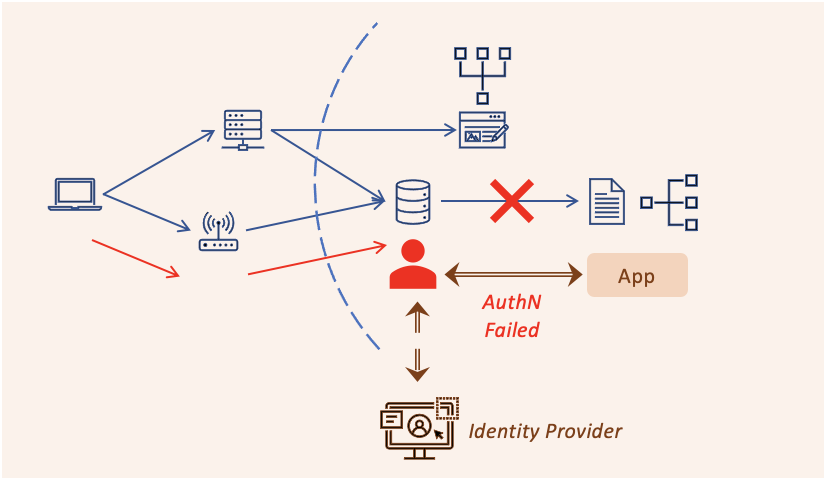}
 \caption{Cyber resilience with authentication layer. After the attacker enters the security perimeter, he still needs to be continuously authenticated before reaching the target. If one of the authentications fails, the attack fails.}
 \label{fig:egauthn}
\end{figure}

\subsubsection*{Cyber Resilience with Authentication Layer:}

Due to the zero-trust principles, network systems are required to continuously monitor the behaviors of the user and adjust access decisions accordingly. After penetrating the system as an insider, the attacker still needs to go through constant authentication processes at each step before reaching the final destination, as illustrated in Figure~\ref{fig:egauthn}. Continuous authentication creates extra effort for the attacker, especially when the system applies strict authentication methods like MFA or Biometric Authentication. The access request would be denied if the attacker fails the authentication, thus the critical asset is protected. Dynamic policy adjustment and continuous authentication help the system recover from existing compromises and prevent further damage to the system.

\begin{figure}[ht!]
 \centering
 \includegraphics[width=0.7\textwidth]{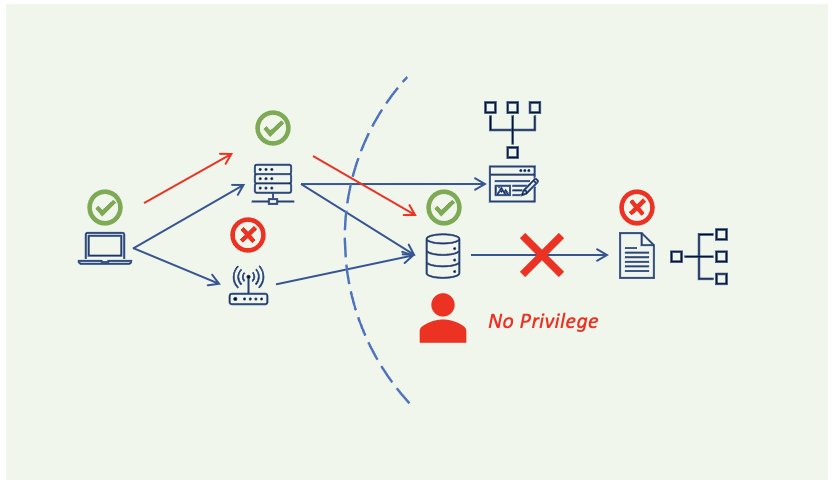}
 \caption{Cyber resilience with authorization layer. After entering the internal network, the attacker still needs explicit privilege in order to access the target. This provides an additional layer of protection for the system.}
 \label{fig:egpri}
\end{figure}

\subsubsection*{Cyber Resilience with Authorization Layer:}

Zero trust security enforces least privilege access as the fundamental principle and ensures that a user is given the minimum levels of permissions to perform legitimate functions at a bounded period. This principle mandates strict policies and permissions for all accounts. With the help of trust evaluation mechanisms and user behavior analytics, the system could downgrade the privilege level of the compromised user account based on abnormal observations, thus stopping the attacker from accessing the critical assets, as in Figure~\ref{fig:egpri}. This layer again provides cyber resilience after the security measures at the traditional perimeter fail.

\begin{figure}[ht!]
 \centering
 \includegraphics[width=0.7\textwidth]{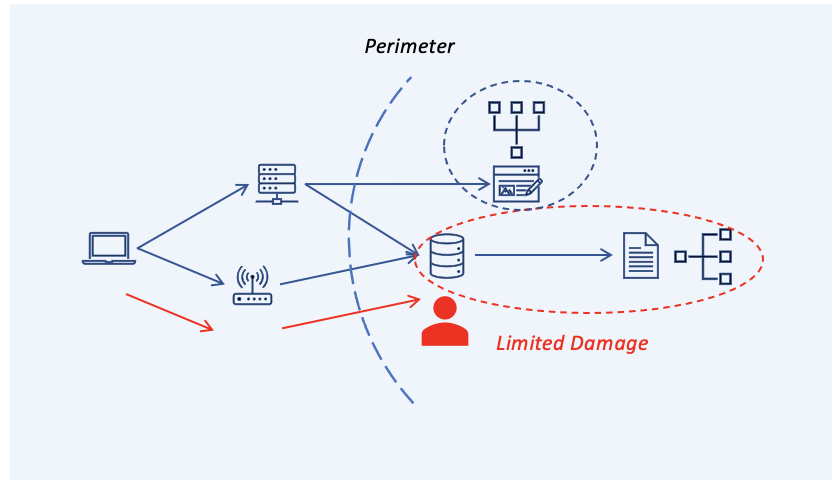}
 \caption{Cyber resilience with network layer. The micro-segmentation at the network layer ensures the potential damage of the attack is limited. Even though the attacker compromises some critical nodes, the network layer protection helps the system withstand the attack and remain functional at other parts of the system.}
 \label{fig:egseg}
\end{figure}

\subsubsection*{Cyber Resilience with Network Layer:}

Finally, micro-segmentation at the network layer limits the attack impact if an external or insider breach does occur. As illustrated in Figure~\ref{fig:egseg}, the protection at the network layer restricts the influence radius of the data breach within only one network segment and prevents further damage to the system. By isolating the workloads, zero trust limits the effect of malicious lateral movement and ensures the security of other parts of the system after the attack. Micro-segmentation offers granular security as network administrators can strengthen and pinpoint security by creating specific policies for critical applications. The segmentation policy can be expressed in terms of workload identities or attributes rather than physical network constructs. Thus, it offers dynamic protection that is adaptive to topology changes, especially in dynamic IoT environments.

\section{Strategic Zero Trust Implementation}
\label{sec:case}

\subsection{A Game Theoretical Approach}

The interdependency between trust evaluation (TE) and policy engine (PE) is critical in zero trust as it helps ensure that trust policies are effective, appropriate, and adaptive to changing threats and risks.  To formalize the zero-trust design, game theory has been a natural and successful framework for modeling the interactions between the system and the potential attacker. Many problems in cybersecurity are fundamentally decision-making problems under complex, uncertain, multi-agent conditions \citep{kamhoua2021game}. The optimal security policy aims to maximize the non-myopic well-being of the system by looking ahead multiple steps into the horizon. The trust score of an agent is evaluated under a given security policy, and the optimal security policy is computed using the Bellman principle based on the trust score of an agent. The game theory equilibrium analysis not only provides a way to quantitatively assess the security posture of the network but also enables a formal methodology to design best-effort zero-trust policies. The optimal security policy aims to maximize the non-myopic well-being of the system by looking ahead multiple steps into the horizon.

In this section, we introduce several works that apply game theory to cybersecurity and achieve cyber resilience with autonomous and proactive zero-trust security responses. The autonomous defense depends on constant monitoring of user activities and makes decisions that are adaptive to online situation awareness. The automation also enables cyber resilience so that the system can adapt to sudden changes in agent behaviors under insider threats. The proposed frameworks consolidate traditional protections with zero-trust security models and provide a guide for a next-generation zero-trust framework.
We hope that by using a game-theoretic approach, the system can achieve automated zero-trust security with better efficiency. Hence, we propose the following strategic zero-trust principle:
\begin{center}
   \textbf{\textit{``Never trust, verify (defense) strategically.''}}
\end{center}

\subsection{Case Studies}

\subsubsection{Proactive Defense Against Insider Threat}

Advanced Persistent Threats (APTs) are a class of emerging threats for cyber-physical systems due to their stealthy, dynamic, and adaptive nature. The deceptive behaviors of the attacker are captured by the multi-stage game of incomplete information, where each player has his private information unknown to the other \citep{huang2020dynamic}. Both players act strategically according to their beliefs about the type of opponent which are formed by multi-stage observation and learning. It is important to develop a proactive and adaptive defense mechanism based on the trust estimation of the user.

Consider a two-player Markov game with incomplete information played on finite stages, $k\in\{0,1,\dots,K\}$. Each player ($i\in\{1,2\}$) has a set of possible types $\Theta_i$, which is private information unrevealed to the opponent. The user's type $\theta_2$ is either adversarial $\theta_2^b$ or legitimate $\theta_2^g$ and the defender's type is either sophisticated $\theta_1^H$ or primitive $\theta_1^L$.  
Based on the current state $x^k\in X^k$, the defender evaluates the trustworthiness of the user (forms a belief $b_1^k\in\Delta(\Theta_2)$) and takes action $a_1^k\in A_1^k$ according to his behavioral strategy $\sigma_1(\cdot)\in \Delta(A_1^k)$. Similarly, the user computes her belief on the defender's sophistication level $b_2^k\in\Delta(\Theta_1)$ and makes a move $a_2^k\in A_2^k$ according to her strategy $\sigma_2(\cdot)\in \Delta(A_2^k)$. The state $x^k$ transfer to the next state $x^{k+1}\in X^{k+1}$ through a known state transition function $f^k$, i.e., $x^{k+1}=f^k(x^k,a_1^k,a_2^k)$. Both players update their beliefs using a Bayesian approach based on observed information,
\begin{align}
    b_i^{k+1}(\theta_j|x^{k+1},\theta_i) = \frac{\Pr[x^{k+1}|\theta_j,x^k,\theta_i]b_i^k(\theta_j|x^k,\theta_i)}{\sum_{\Bar{\theta}_j\in \Theta_j}\Pr[x^{k+1}|\Bar{\theta}_j,x^k,\theta_i]b_i^k(\Bar{\theta}_j|x^k,\theta_i)}
    \qquad i,j\in\{1,2\}, j\neq i.
    \label{eq:bayes}
\end{align}
With the conditional independence of the player's strategies $\sigma_1^k$ and $\sigma_2^k$,
\begin{align}
    \Pr[x^{k+1}|\theta_j,x^k,\theta_i]= \sum_{a_1^k\in A_1^k,a_2^k\in A_2^k} f^k(x^k,a_1^k,a_2^k) \sigma_1^k(a_1^k|x^k,\theta_1)
    \sigma_2^k(a_2^k|x^k,\theta_2).
\end{align}
Under this setting, the trust score of the user at stage $k$ can be interpreted as the defender's belief that the user is legitimate to the system, i.e., $TS^{k}=b_1^{k}(\theta_2=\theta^g)$.

\begin{figure}[!t]
 \centering
 \includegraphics[width=0.9\textwidth]{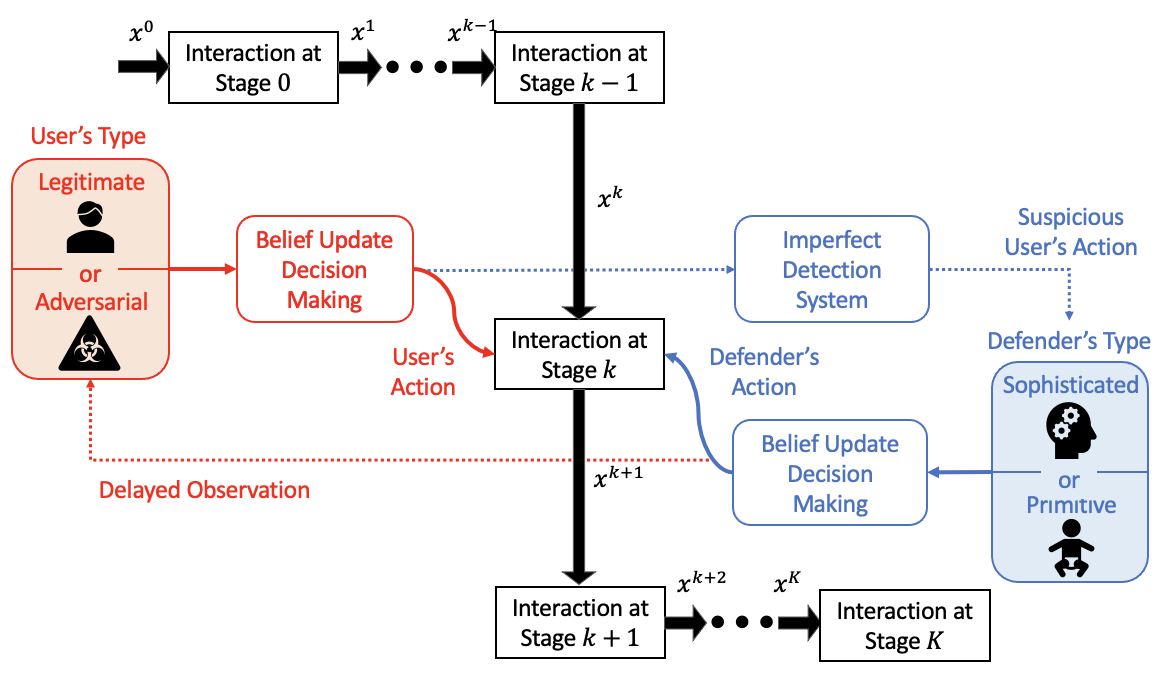}
 \caption{An illustration of proactive defense against multi-stage APT attacks. We denote the user, the defender, and the system states in red, blue, and black, respectively. The defender interacts with the user from stage $0$ to stage $K$ in a sequence where the output state of stage $k-1$ becomes the input state of stage $k$. At each stage, the user observes the defender’s actions at previous stages, forms a belief in the defender’s type, and takes an action. At the same time, the defender makes decisions based on the output of an imperfect detection system.}
 \label{fig:apt}
\end{figure}

Both players at each stage should optimize their expected cumulative utilities concerning the updated beliefs at the future stages, which leads to the Perfect Bayesian Nash Equilibrium (PBNE). A perfect Bayesian Nash equilibrium consists of two conditions: belief consistency and sequential rationality. The belief consistency emphasizes that when strategic players make long-term decisions, they have to consider the impact of their actions on their opponent’s beliefs at future stages. Sequential rationality guarantees that unilateral deviations from the equilibrium at any state do not benefit the deviating player. \citep{huang2020dynamic} provides a computational algorithm to find the relaxed $\epsilon$-PBNE.

\begin{figure}[!t]
 \centering
 \includegraphics[width=0.9\textwidth]{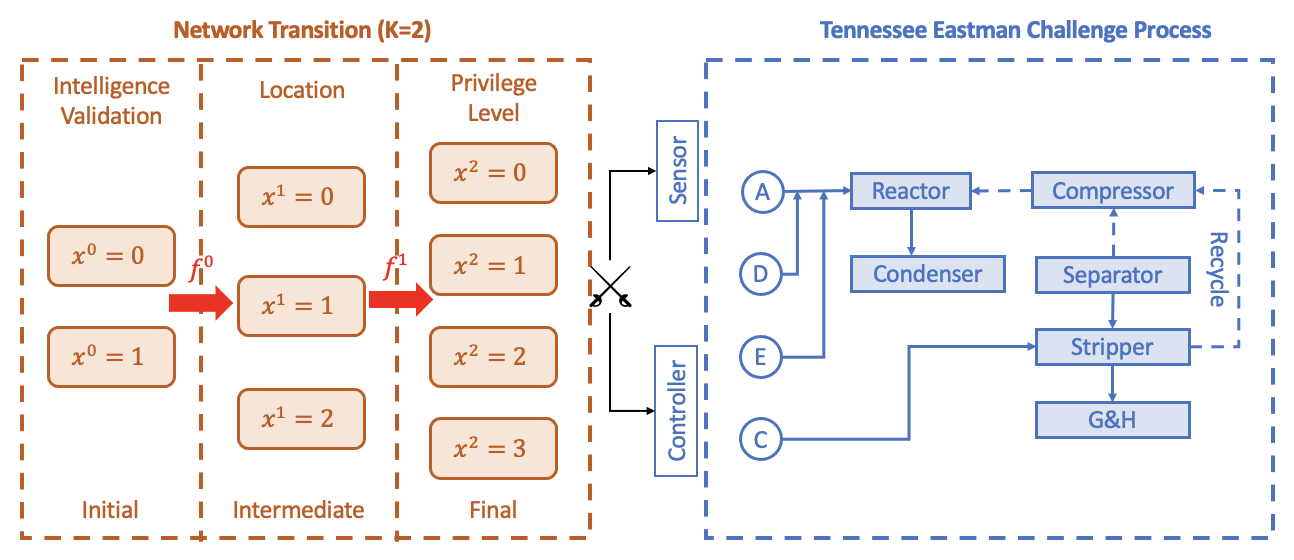}
 \caption{The cyber state transition and the physical attack on Tennessee Eastman process.}
 \label{fig:tennessee}
\end{figure}

Figure~\ref{fig:tennessee} illustrates the Tennessee Eastman process as a benchmark case study of industrial control systems. 
The initial stage $k=0$ contains two states to show whether the reconnaissance is effectual $x^0=1$ or not $x^0=0$. The defender could choose to conduct different levels of security training to increase employees’ security awareness and protect them from web phishing. The user (potential attacker) can send emails with non-executable attachments and shortened URLs to the accounts of entry-level employees, managers, or avatars. The state at the intermediate stage $k=1$ can be interpreted as the location of the user. The user can choose to escalate his privileges or perform no operation. On the other hand, the defender decides whether to permit the privilege escalation or not based on his assessment of the trustworthiness of the user. The decision is based on the trust estimation of the user via Bayesian updates in \eqref{eq:bayes}. At the final stage $k=2$, the state represents the privilege levels of the user. Based on the types of both players $(\theta_1,\theta_2)$, their action choices $(a_1^k,a_2^k)$, and the underlying state $x^k$, we estimate the utilities by considering command injection attacks at supervisory control and data acquisition (SCADA) systems in Tennessee Eastman process.

\begin{figure}[!t]
 \centering
 \includegraphics[width=0.7\textwidth]{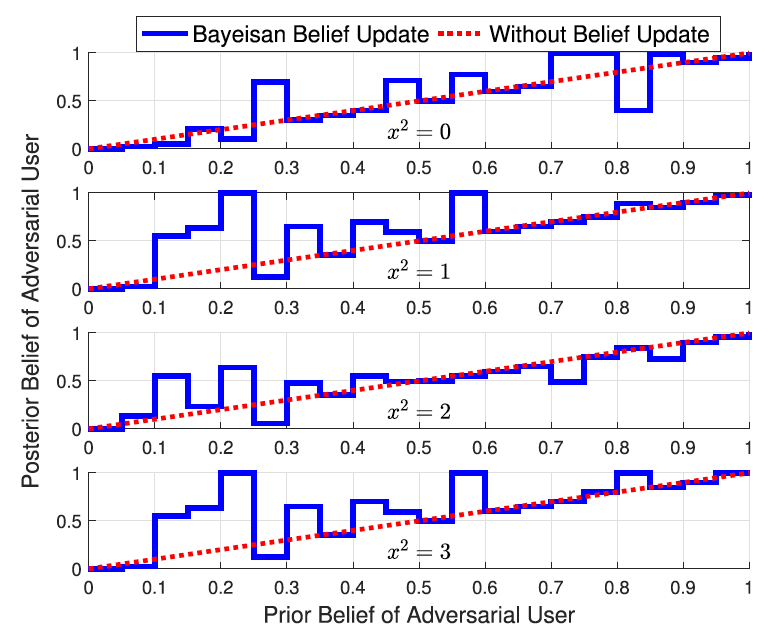}
 \caption{The defender’s prior and posterior beliefs of the user being adversarial.}
 \label{fig:beliefupdate}
\end{figure}

Experimental results illustrate that the private types of players would affect their policies and utilities under different information structures. It is important to establish a reliable trust assessment mechanism so that the defense policy remains effective. Besides, the Bayesian belief update leads to a more accurate estimate of users’ types. By considering the prior knowledge, side evidence, and observed strategies, the defender effectively discovers the true type of user. As illustrated in Figure~\ref{fig:beliefupdate}, Bayesian update helps to correct the inaccurate static prior belief and allows the defense strategies to remain effective. One interesting observation is that the defender benefits from introducing defensive deception, e.g., a primitive defender can disguise himself as a sophisticated one to confuse the attacker. Interested readers can refer to \citep{pawlick2021game} for more examples and applications.

\subsubsection{Strategic Trust in Cloud-enabled Cyber-physical Systems} 

\begin{figure}[!t]
 \centering
 \includegraphics[width=0.6\textwidth]{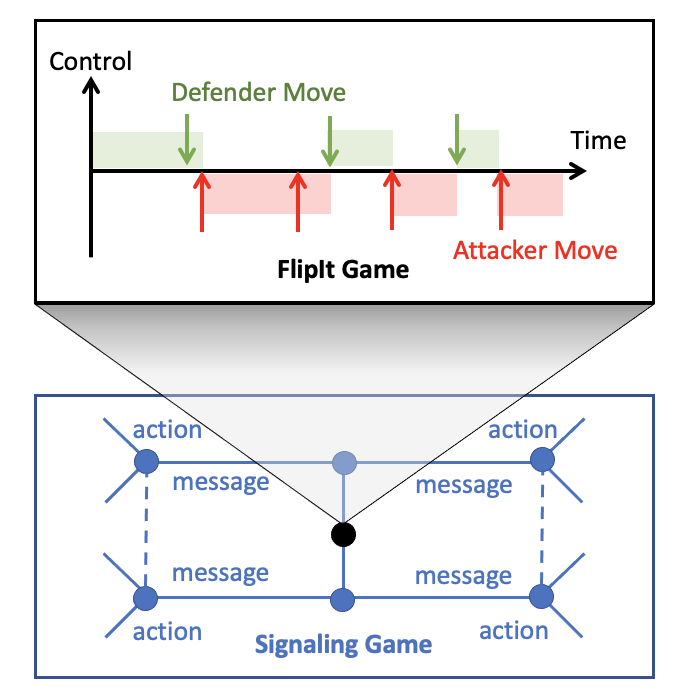}
 \caption{Conceptual model of the combined cloud control game. The attacker ($\mathcal{A}$) and Defender ($\mathcal{D}$) play the FlipIt game for control of the cloud. Then, the winner sends a command to the device (R) in the signaling game.}
 \label{fig:metagame}
\end{figure}

Zero-trust security is important in cyber-physical systems (CPS). The integrated physical and computational capabilities in CPS have created a new generation of systems. However, the heterogeneous components in such systems enlarge the attack surface and leave multiple vulnerabilities that the attacker could explore. Trust in devices and signals becomes a critical factor in decision-making. Each device must decide whether to trust other components that may be compromised. In \citep{pawlick2017strategic}, we use a game-theoretic approach to evaluate the trust of the cloud service based on the potential consequence of the trust. The trust is based on the game-theoretic assessment of the attack and defense strategies. The trust game and the defense games are consolidated into a meta-game framework, as shown in Figure~\ref{fig:metagame}, to jointly design trust and defense mechanisms to achieve a zero-trust architecture.

Each agent in a CPS must decide whether to trust signals from the vulnerable cloud that may be compromised. We refer to this trust as strategic trust, which is a positive belief about the perceived reliability of, dependability of, and confidence in the signal. The trust of the signal is estimated using the FlipIt game $\mathbf{G_F}$ \citep{van2013flipit}, in which an attacker ($\mathcal{A}$) and a defender($\mathcal{D}$) are competing to control the cloud service. The players choose how often to move based on their incentives to control the cloud and the cost that they pay to capture it. Nash equilibrium of $\mathbf{G_F}$ reveals the probability with which the cloud is compromised in the steady state of the game. In equilibrium, the attacker $\mathcal{A}$ controls the cloud service with probability $p\in[0,1]$. Then, the trustworthiness of the cloud signal can be written as $TS^0=1-p\in[0,1]$, which serves as the prior belief in the following signaling game between the cloud and the device ($\mathcal{R}$). 

In the following signaling game $\mathbf{G_S}$, the sender may be an attacker with type $\theta_A$ or a defender with type $\theta_D$. The device  $\mathcal{R}$ does who controls the cloud (the type of the sender). Thus $\mathcal{R}$ needs to decide whether to trust the received command (message $m$) based on his belief (trust) about the unknown type of the sender. This posterior belief is computed in a Bayesian fashion by considering the prior trust $TS^0$ from the FlipIt game and the sender's strategies under the equilibrium of $\mathbf{G_S}$.

The trust-based decision is vital, especially in CPSs related to critical infrastructures, vehicle networks, personal health products, etc. The consequence of wrongly trusting an attacker could be severe as it could involve public operation or human safety. The Gestalt Nash Equilibrium (GNE) \citep{chen2018security} of the overall macro-game $(\mathbf{G_F},\mathbf{G_S})$ characterizes the steady state and risk assessment in CPS. It helps to study the strategic trust of vulnerable network components and contributes to the design of zero-trust security in CPS.

\subsubsection{Strategic Zero-trust Authentication for 5G Networks}

The recent trends in remote work and bring your own device (BYOD) exacerbate the need for a trustworthy 5G network system to provide proper authentication mechanisms for external connections from unknown or uncertain environments. 5G network can be divided into a public network that is open to any connection and a private network that is fully controlled by enterprise or government. Although the private network is composed of trusted devices, it still needs to exchange information with the public network due to external access or cloud services. If a remote agent from a public network is trying to access a critical node in the private network, it is essential to evaluate the trust of the agent before granting access. An attacker may utilize the stored credentials in the already compromised nodes and penetrate the system to reach some critical assets. To prevent such lateral movements of the potential attacker, the zero-trust system should strategically make authentication decisions based on the trust evaluation of the agent. Since the system cannot fully observe the true identity of the agent, the system needs to estimate the trustworthiness of the agent based on past observations and interactions. 
In \citep{ge2022mufaza}, we have proposed a formal zero-trust security framework called
GAZETA, building on Markov games with one-sided information, to capture the dynamic interactions between a potential attacker and defender.

\begin{figure}[!t]
 \centering
 \includegraphics[width=0.9\textwidth]{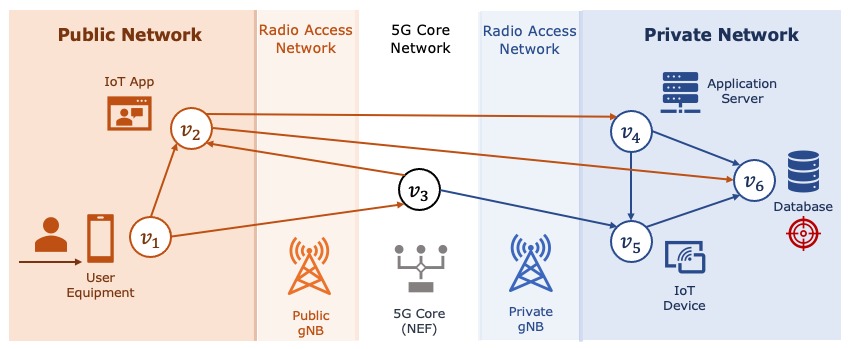}
 \caption{Example of an authentication graph in a hybrid 5G-enabled IoT network. The agent initially logs into the system using the user’s equipment. The agent uses the authentication graph to try to reach the target node, i.e., the database.  The defender needs to evaluate the trust of the agent and apply a strategic authentication policy accordingly.}
 \label{fig:5gcase}
\end{figure}

The game is played on the authentication graph in Figure~\ref{fig:5gcase}, which is an agent-centric directed graph $G=\langle V,E\rangle$ to describe the network-level authentication activities and reachability of the agent. By monitoring the logon events or processes that search for user credentials, each node $v\in V$ in the network can be labeled by an indicator function $L(v)=\mathbbm{1}$(node $v$ has been visited). The authentication graph and labeling function over all nodes consist of the game state at each stage $k=\{1,2,\dots,K\}$. A binary type set $\Theta=\{0,1\}$ is considered for the agent, where $\theta=0$ is a legitimate user and $\theta=1$ represents a malicious attacker. The TS of the agent is defined as the belief the agent is legitimate at stage $k$, i.e., $TS^k=b^k(\theta=0)$. At each stage, the agent attempts to access the next node using the stored credentials in the visited nodes. The defender aims to validate the authentication of the agent and reject potential lateral movement from malicious attackers. The identity validation may require alternative authentications (e.g., Multi-factor
Authentication (MFA)), which would create difficulty for the attacker with only stored credentials. However, this additional authentication process is resource- and time-consuming. The defender needs to discover attacks while minimizing resource consumption for the sake of balancing system performance and security. The game is played until the finite game horizon $K$, which represents the maximum time span for the agent to exist in the network without credential renewal. The attacker will lose his foothold in the network if he cannot reach the target within $K$ steps.

\begin{figure}[!t]
 \centering
 \includegraphics[width=0.7\textwidth]{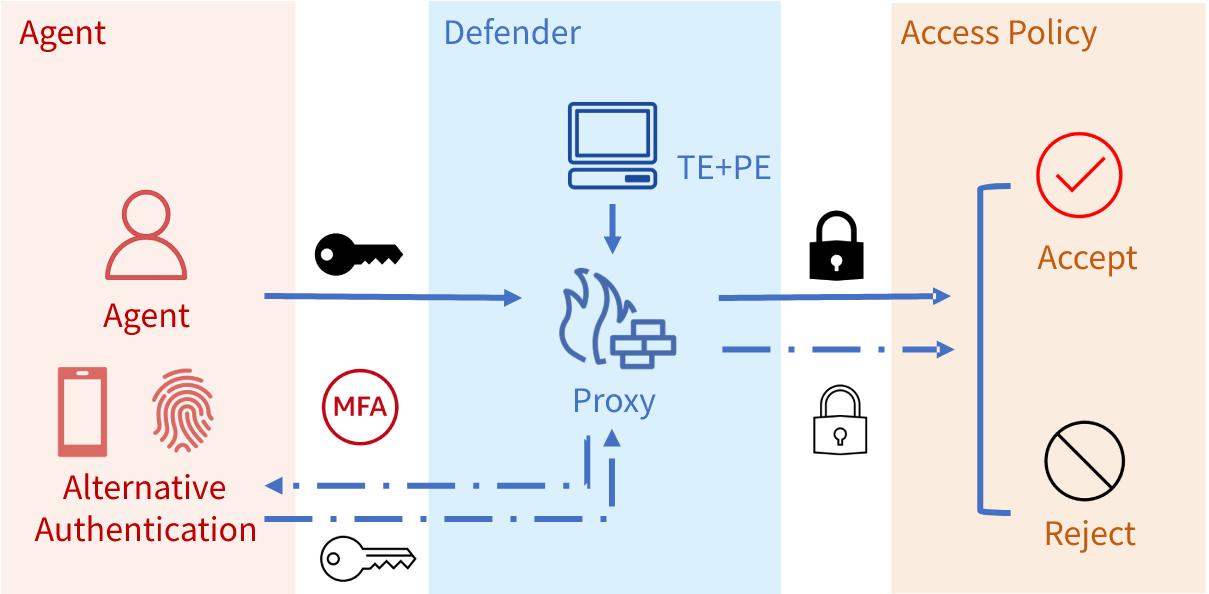}
 \caption{Illustration of the authentication process. The defender decides whether to request authentication validation before granting access.}
 \label{fig:authnprocess}
\end{figure}

The game commits to one-sided $\epsilon$-PBNE, which is a refined $\epsilon$-PBNE when there is only one player (the defender) who cannot fully observe the type of the opponent (attacker). To improve cyber robustness and resilience to unexpected environmental changes, \citep{ge2022mufaza} proposed a moving-horizon algorithm to approximately compute the optimal strategies online. The online zero-trust defense enables the system to integrate side evidence into trust evaluation, identify the attacker quicker, and eliminate the potential risks of a persistent attack.

Experiments on the 5G network case study in Figure~\ref{fig:5gcase} show that the strategic zero-trust defense outperforms other access control methods by providing the shortest access time for the benign user and the longest access time for the malicious attacker. It has been corroborated that the defense mechanism can efficiently correct erroneous prior trust and provide a more reliable trust evaluation. With moving-horizon computation, GAZETA has created a robust security model that can adapt to unexpected environmental changes and account takeover. Experimental results show that the online defense scheme can quickly discover the account takeover and deter the lateral movement of the attacker successfully. The proposed mechanism provides a guide for the development of next-generation secure and resilient 5G zero-trust networks.

\section{Conclusion}
\label{sec:conclusion}

Zero-trust is an emerging security concept based on a strict verification process. This chapter draws attention to the cyber resilience within the zero-trust model. We first introduced the evolution from traditional perimeter-based security to zero trust and discuss the differences between them. We provide the essential elements of trust evaluation (TE) design and introduce the trust-layer perspective mitigation in the policy engine (PE). The interdependence between TE and PE results in an efficient and adaptive security policy that contributes to cyber resilience in every stage of an attack. We used a running example to illustrate what zero trust can do for cyber resilience. To model the zero-trust framework, we utilize a game-theoretic approach to achieve the proposed strategic zero-trust principle. Three case studies are provided to illustrate how to apply game theory to cybersecurity and achieve cyber resilience with autonomous and proactive zero-trust security responses.  The proposed mechanisms consolidate traditional protections with zero-trust security models and provide a guide for the development of next-generation secure and resilient zero-trust networks.

\bibliographystyle{plainnat}
\bibliography{wiley}%

\end{document}